\documentclass[aps,superscriptaddress,amssymb,showpacs,prb,twocolumn]{revtex4}
\usepackage{graphicx}
\begin{document}

\title{Monte Carlo studies of quantum and classical annealing on a double-well}

\author{Lorenzo Stella}
\affiliation{International School for Advanced Studies (SISSA), and INFM
Democritos National Simulation Center, Via Beirut 2-4, I-34014 Trieste, Italy}
\author{Giuseppe E. Santoro}
\affiliation{International School for Advanced Studies (SISSA), and INFM
Democritos National Simulation Center, Via Beirut 2-4, I-34014 Trieste, Italy}
\affiliation{International Centre for Theoretical Physics
(ICTP), P.O.Box 586, I-34014 Trieste, Italy}
\author{Erio Tosatti}
\affiliation{International School for Advanced Studies (SISSA), and INFM
Democritos National Simulation Center, Via Beirut 2-4, I-34014 Trieste, Italy}
\affiliation{International Centre for Theoretical Physics
(ICTP), P.O.Box 586, I-34014 Trieste, Italy}

\date{\today}

\begin{abstract}
We present results for a variety of Monte Carlo annealing approaches, both 
classical and quantum, benchmarked against one another for the textbook optimization 
exercise of a simple one-dimensional double-well. In classical (thermal) annealing, 
the dependence upon the move chosen in a Metropolis scheme is studied and 
correlated with the spectrum of the associated Markov transition matrix.
In quantum annealing, the Path-Integral Monte Carlo approach is found to
yield non-trivial sampling difficulties associated with the tunneling 
between the two wells. The choice of fictitious quantum kinetic energy is
also addressed. We find that a ``relativistic'' kinetic energy form, 
leading to a higher probability of long real space jumps, can be considerably
more effective than the standard one.
\end{abstract}

\maketitle

\section{Introduction}\label{Intro:sec}
%
Solving difficult classical combinatorial problems by adiabatically controlling 
a fictitious quantum Hamiltonian is a fascinating route to quantum computation \cite{Vazirani}, 
which is alternative to the main-stream approach \cite{QC}. 
The quantum annealing (QA) strategy \cite{Finnila:article, QA_Kolkata:proceeding} is precisely based 
on the adiabatic switching off, as a function of time, of strong fictitious quantum 
fluctuations, suitably introduced by an extra kinetic term in the otherwise 
classical Hamiltonian under consideration.

A number of authors have by now applied QA to attack a variety of optimization problems.
Some amount of success was obtained in several cases, notably in the folding of 
off-lattice polymer models \cite{Berne1:article,Berne2:article}, in the 2D random 
Ising model annealing\cite{Santoro:science,Martonak:ising}, in the Lennard-Jones 
clusters optimization\cite{Berne3:article,Gregor-Car:article}, and in the Traveling 
Salesman Problem \cite{Martonak:TSP}.

Despite these successes -- mostly obtained by Path-Integral Monte Carlo (PIMC) implementations
of QA -- there is no general theory addressing and even less predicting the 
performance of a QA algorithm. Different optimization problems are characterized
by different ``energy landscapes'' of barriers and valleys about which 
little is known but which clearly influence the annealing process\cite{TSP_landscape:article}.
Success of QA crucially depends on the efficiency with which the chosen kinetic 
energy and associated fictitious quantum fluctuations explores and influences 
these effective energy landscape.
This is a uncomfortable situation, in view of the fact that it is \emph{a priori} not 
obvious or guaranteed that a QA approach should do any better than, for instance, Classical
Simulated Annealing (CA). Indeed, for the interesting case of Boolean 
Satisfiability problems \cite{TCS:article} -- more precisely, a prototypical 
NP-complete problem such as 3-SAT -- a recent study has shown
that PIMC-QA performs definitely {\em worse} than simple CA \cite{demian_QA:article}.

In a previous paper \cite{Stella_simple:article}, henceforth referred to as I, 
we compared classical and quantum annealing approaches in their performance
to optimize the simplest potential landscape $V(x)$, namely the one-dimensional 
potential double-well. Classical annealing was implemented by means of a 
Fokker-Planck (FP) equation with a time-dependent (decreasing) temperature 
$T(t)$. Quantum annealing was performed through the Schr\"odinger's equation, 
both in real and in imaginary time, describing a particle with a time-dependent 
(decreasing) inverse mass $m^{-1}(t)$ in the double-well potential $V(x)$. Given 
the textbook simplicity of the problem, it was possible to correlate in detail 
the features of the landscape (position and shape of valleys, barrier height)
with the outcome of the annealing process. One saw for instance that 
Fokker-Plank CA is sensitive only to the ratio $\Delta_V/B$ of the 
energy splitting $\Delta_V$ between the two wells, or valleys, and the barrier 
height $B$, whereas on the contrary Schr\"odinger QA is sensitive to the 
Landau-Zener tunneling gap, and hence, among other things, to the barrier width.

Unfortunately, such a direct approach is applicable, for practical purposes, 
only to insignificantly small sized optimization problems. To appreciate the 
difficulty, it is enough to consider that the number of possible configurations, 
and hence the Hilbert space size, of a small 32$\times$32 square lattice 
Ising model is $2^{1024} \sim 10^{308}$, an astronomically large number which 
forbids any direct approach based on deterministic state evolution. Hard 
instances of typical optimization problems involve an even larger number 
of variables. Because of that, alternative strategies, based on Monte Carlo
sampling, are mandatory.
That notwithstanding, the various deterministic types of dynamics, Fokker-Planck, 
Schr\"odinger, or Monte Carlo dynamics, classical or quantum, are not at all 
equivalent. There is in particular no relationship between the (physical) time 
appearing in the Fokker-Planck or in the Schr\"odinger equation, 
and the corresponding classical or quantum Monte Carlo time-step. 
Furthermore, {\em many} different types of Monte Carlo (MC) dynamics are possible.
While sampling the same equilibrium probability distribution and hence providing the
same equilibrium averages \cite{LandauBinder:book}, they will generally 
perform very differently in an out-of-equilibrium situation. In order to
appreciate that, given the large variability of the possible MC outcomes, 
it is once again wise to concentrate attention on landscapes which are well under control. 

In this paper we implement and investigate different Monte Carlo annealing 
strategies, both classical and quantum, to the very same optimization problem -- 
the one-dimensional double-well -- which was studied in I by deterministic 
(Fokker-Planck and Schr\"odinger) annealing approaches.  
We will first show (Sec.~\ref{CA_intro:sec} that the choice of the proposed 
move in a Metropolis MC scheme strongly influences the annealing performance. 
Such a strong influence of the move is seen to be correlated with the instantaneous 
spectrum of the Markov transition matrix associated to the MC scheme, as shown 
in detail in Sec.~\ref{Spectral:subsec}. Next, we will turn to  Quantum MC, 
specifically to Path-Integral Monte Carlo (PIMC), and show how the required 
tunneling dynamics is highly non-trivial. The outcome of an allegedly
state-of-the-art PIMC annealing of a simple double-well can even be very 
unsatisfactory, due to sampling difficulties of instanton events 
(see Secs.~\ref{PIMC:sec}, \ref{results1:subsec}, \ref{results2:subsec}).
Finally, we will analyze the choice of kinetic energy in the QA Hamiltonian,
by testing the effectiveness of an alternative {\it relativistic} kind
of dispersion, which proves much more effective than the usual non-relativistic 
choice (Sec.~\ref{results3:subsec}). 
Section \ref{discussion:sec} will finally contain a summary of the main results 
some discussion. Technical details on the spectral analysis of the
classical Markov process, and on the generalization of the bisection algorithm 
of interest for the PIMC study of Sec.~\ref{results3:subsec}, are relegated to the two Appendices.

\section{Classical Monte Carlo annealing of a double-well.} \label{CA_intro:sec}
%
Let us start with the double-well one-dimensional potential on which we will test the different
classical and quantum Monte Carlo (MC) annealing schemes presented in this paper. 
We assume $V(x)$ to describe two generally asymmetric wells of the form:
\begin{equation} \label{Vasym:eqn}
V(x) =
\left\{ \begin{array}{c}
         V_0 \frac{(x^2-a_+^2)^2}{a_+^4} + \alpha x
             \hspace{10mm} \mbox{for} \; x\ge0 \\
         V_0 \frac{(x^2-a_-^2)^2}{a_-^4} + \alpha x
             \hspace{10mm} \mbox{for} \; x < 0 \\
       \end{array}
\right. \;,
\end{equation}
where $a_+\ge a_-$ (both positive), and the linear term $\alpha x$,
with $\alpha \ge 0$, splits the degeneracy of the two minima.
(The discontinuity in the second derivative at the origin is of no consequence
in our discussion.)
The optimal value of the potential is obviously $E_{opt}=V(x_-)$. 
To linear order in the small parameter $\alpha$, the two minima are
located at $x_{\pm}=\pm a_{\pm} - \alpha a_{\pm}^2/(8V_0)$,
and the splitting between the two minima is 
$\Delta_V = V(x_+)-V(x_-)=\alpha (a_{+}+a_{-})$, while the second derivative of the
potential at the two minima, to lowest order in $\alpha$, is given by
$V^{\prime\prime}(x=x_{\pm}) = {8V_0}/{a_{\pm}^2}$.
A ``symmetric'' double-well potential is recovered for $a_+=a_-$. 

\begin{figure}
\begin{center}
\includegraphics[width=8cm]{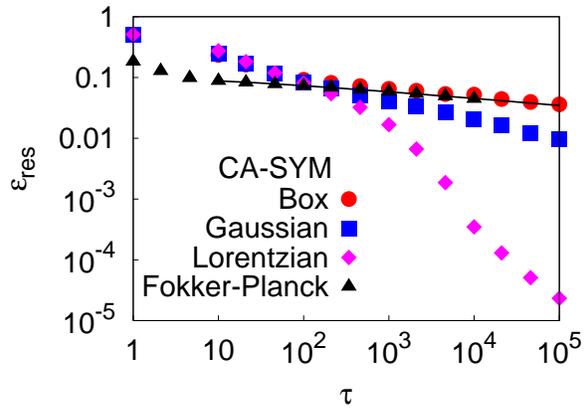}
\end{center}
\caption{Classical annealing. Comparison between Fokker-Planck (deterministic) annealing (triangles), 
  and several different Metropolis Monte Carlo annealings which differ only by the
  choice of the proposed move: Box, Gaussian, or Lorentzian (see text).
  The potential is a symmetric double-well of Eq.~\protect\ref{Vasym:eqn}
  with $a_-=a_+=1$, $V_0=1$, and a linear term $\alpha=0.1$.  
  The initial condition was $T_0=V_0=1$. 
  The solid line through the points is a fit with a power-law of leading exponent 
  $\tau^{-\Delta_V/B}$ times logarithmic corrections, see I.
  Clearly, the Fokker-Planck and the Monte Carlo annealing ``times'' are not related (see text).}
\label{FP_vs_MC_SYM_fdt:fig}
\end{figure}

Starting with some initial state or distribution, the key quantity 
under inspection will be the average {\rm residual energy} 
$\epsilon_{res}(\tau)=\bar{E}_{pot}(\tau)-E_{opt}$ after annealing for a certain time 
$\tau$, where $\bar{E}_{pot}(\tau)$ is the average final potential energy. 
In the classical MC case, annealing is performed by reducing the temperature 
$T(t)=T_0(1-t/\tau)$. 
As shown in I (Ref.~\onlinecite{Stella_simple:article}), the results of a Fokker-Planck annealing are
roughly independent of the asymmetry of the potential, and only controlled by the
ratio $\Delta_V/B$ of the splitting $\Delta_V$ between the two minima and
the barrier height $B=V_0-V(x_+)$. 
In Fig.~\ref{FP_vs_MC_SYM_fdt:fig} we present the results
of a standard Metropolis MC classical annealings for a symmetric 
($a_-=a_+=1$, $V_0=1$) double-well potential, 
where a linear term $\alpha x$ with $\alpha=0.1$ provides a splitting 
between the two minima of order $\Delta_V\approx 0.2$.
Entirely similar results (not shown) are obtained for asymmetric double-wells.
The solid triangles represent the results of exact integration of Fokker-Planck's (FP) 
equation, via a fourth-order Runge-Kutta method, with a linear annealing schedule for the
temperature $T(t)=T_0(1-t/\tau)$, as reported in I. 
%
%
The circles, squares, and diamonds represent the results of three different classical 
Metropolis MC annealings -- clearly showing very different behaviors --, differing uniquely 
in the choice of the proposed move.
To clarify this point, we recall that the Markov process associated with a Metropolis MC
is fully specified by a transition probability $W(x^\prime|x)$, to move 
from $x$ to $x^\prime$ in the time step $\Delta t$. W is the product of two factors: 
\begin{equation} \label{Metropolis_T:eqn}
W(x^\prime|x) = A(x^\prime|x) P(x^\prime|x) \hspace{4mm} \mbox{for} \; x^\prime\ne x \;, 
\end{equation}
where $P(x^\prime|x)$ is the probability of {\em proposing} a move from $x$ to $x^\prime$, while 
\begin{equation}\label{rejection} 
A(x^\prime|x) = {\rm Min} 
\left\{1,{P(x|x^\prime)  \over P(x^\prime|x)} e^{-[E(x^\prime)-E(x)]/k_BT} \right\} \;,
\end{equation}
is the probability of {\em accepting} the proposed move. 
The various results shown in Fig.~\ref{FP_vs_MC_SYM_fdt:fig}
differ in the choice of $P(x^\prime|x)$, i.e.:\\
1) (Box, circles) the new proposed position $x^\prime$ is uniformly distributed inside a box of 
length $2\sigma$ centered around $x$, i.e.,
\begin{equation} \label{P_transition_matrix_box:eqn}
P(x^\prime|x)=P_{\rm Box}(x^\prime|x)=\frac{1}{2\sigma}\theta(\sigma-|x^\prime-x|) \;;
\end{equation}
2) (Gaussian, squares) the new position $x^\prime$ is distributed according to a Gaussian of
width $\sigma$ centered around $x$, i.e.,
\begin{equation} \label{P_transition_matrix_gau:eqn}
P(x^\prime|x)=P_{\rm Gau}(x^\prime|x)=\frac{1}{\sqrt{2\,\pi}\sigma} 
      e^{-(x^\prime-x)^2/(2\sigma^2)} \;;
\end{equation}
3) (Lorentzian, diamonds) the new position $x^\prime$ is distributed according to a Lorentzian of
width $\sigma$ centered around $x$, i.e.,
\begin{equation} \label{P_transition_matrix_lor:eqn}
P(x^\prime|x)=P_{\rm Lor}(x^\prime|x)= \frac{1}{\pi}\frac{\sigma}{(x^\prime-x)^2 + \sigma^2} \;.
\end{equation}
In all three cases, the value of $\sigma$, which controls the move range, was
decreased with temperature according to $\sigma=\sigma_0\sqrt{T/T_0}$, 
with $\sigma_0=2.0$, for the Box and the Gaussian case, and $\sigma_0=2.9$ for the
Lorentzian one. This choice was made empirically by monitoring the average acceptance 
and the root mean square displacement, $\sqrt{\langle (x_{i+1} - x_{i})^2 \rangle}$,
of the MC process.

As evident from Fig.~\ref{FP_vs_MC_SYM_fdt:fig}, the results
of the various choices of proposal moves differ in a substantial way. 
While the Box choice leads to a residual energy very close to the deterministic 
Fokker-Planck result, the Gaussian choice, and even more so the Lorentzian one, lead to
residual energies which decrease faster with the annealing time $\tau$. 

The reason for the non uniqueness of the result of a classical MC annealing is not
difficult to grasp. The Metropolis algorithm guarantees that the expectation values
of a system at equilibrium are correctly reproduced by a MC time average, 
after a suitable equilibration time $\tau_{eq}$. 
The equilibration time $\tau_{eq}$ depends on the quantity one is interested, 
and, more crucially for our scope, on the type of proposed moves, i.e., on the $P(x^\prime|x)$. 
For equilibrium MC simulations, this issue is of little concern: As long as 
one checks that the stochastic process has reached a stationary condition, the
results are guaranteed to be correct. In an annealing simulations, however, 
(or, in that respect, in any out-of-equilibrium evolution),
the choice of the $P(x^\prime|x)$, determining the instantaneous equilibration time of 
the process, strongly influences the overall performance of the algorithm. 

Evidently, both the Gaussian and the Lorentzian trial moves allow the attainment
of residual 
energies $\epsilon_{res}(\tau)$ much below that of the Box move (and of the FP dynamics).
The Lorentzian case is particularly impressive.  
If we fit the asymptotic part of the residual energy data by
$\epsilon_{res}(\tau) \propto \tau^{-\Omega_{CA}}$, the 
annealing exponent $\Omega_{CA}$ turns out to be $\sim 0.31$ for the Gaussian case,
and $\sim 1.0$ for the Lorentzian one, i.e. definitely much larger than what the Huse and
Fisher theory \cite{Huse-Fisher:article} predicts for the Fokker-Planck case, namely
$\Omega_{CA}=\Delta_V/B\sim 0.2$.

The long Lorentzian tails clearly imply an abundance of moves operating
{\em long jumps}, which are beneficial to the annealing even at small temperatures $T$.
Ref.~\onlinecite{FastSA,GenSA} indeed have reported a genuine improvement of the sampling by 
making use of the Lorentzian distribution, instead of the Gaussian, for the 
Thompson problem (finding the equilibrium distribution of $N$ charged particles on a sphere) \cite{GenSA}, 
and in optimizing the structure of a cluster made of $N$ Ni atoms \cite{GenSA}.
In order to analyze these issues in more quantitative detail, we carried out a spectral 
analysis of the transition matrices $W(x^\prime|x)$ corresponding to the
different proposal move schemes. 
In the following section we briefly recall the crucial steps behind this approach.
 
\subsection{Spectral analysis of the Markov process in Monte Carlo} \label{Spectral:subsec}
%
Every discrete-time Markov process can be defined by means of an equation of the form:
\begin{equation} \label{markovprocess}
P(i,t+\Delta t) = \sum_{j}\,W(i|j)\,P(j,t) \;,
\end{equation}
where, for simplicity of notation, we have assumed that the possible states of the system, 
labeled $i$ and $j$, are also discrete (we will indeed reduce our continuous problem to
a discrete one in the following numerical calculations, see below).
$P(i,t)$ is here the probability of finding the system in state $i$ at time $t$, and 
$W(i|j)$ is the so-called \emph{Transition Matrix}, i.e., the transition 
probability for going from state $j$ to state $i$ during the time step $\Delta t$.
The transition matrix $W(i|j)$ (assumed to be time-independent) is a stochastic 
matrix \cite{VanKampen:book}, in the sense that $W(i|j)\ge 0$ and $\sum_i W(i|j)=1$. 
In our particular case, the stochastic process behind the Metropolis 
sampling is given by Eq.~\ref{Metropolis_T:eqn}, so that $W(i|j)$ also fulfills 
the detailed balance condition \cite{LandauBinder:book}.
The idea underlying the spectral analysis of $W(i|j)$ is very simple. 
For a fixed temperature $T$, the Markov process admits a spectral analysis
in much the same spirit in which one analyzes a time-dependent Schr\"odinger 
equation via the spectrum of the associated Hamiltonian. 
Indeed, in continuous time -- i.e., for $\Delta t\to 0$ -- the result is quite simple to state. 
A similar analysis can be carried out for the original discrete-time Markov chain, 
see Ref.~\onlinecite{VanKampen:book}.
As explained there, from $W(i|j)$ one can define a closely
related matrix $\bar{W}(i|j)$ appearing in the short-time expansion,
$W(i|j)=\delta_{i,j}+\Delta t \,\bar{W}(i|j)$, of the exponential
map: $W=e^{\Delta t\,\bar{W}}$;
in turn, $-\bar{W}$ possesses a set of right eigenvectors $\left\{ P_k(i) \right\}$
with $P_0(i)=P^{(eq)}(i)$ (the equilibrium distribution), and the corresponding eigenvalues
$\left\{ \bar{\lambda}_k \right\}$ are such that $\bar{\lambda}_0=0$, $\bar{\lambda}_k>0$ $\forall k>0$.
In terms of the spectrum of $\bar{W}$, one can formally write the general
solution of the time-continuum ($\Delta t\to 0$) limit of Eq.~(\ref{markovprocess}) as
\cite{VanKampen:book}:
\begin{equation} \label{spectral_decomposition:eqn}
P(i,t) = P^{(eq)}(i) + \sum_{k>0}\,a_k\, e^{-\bar{\lambda}_k \, t}\,P_k(i) \;,
\end{equation}
where $a_k$ are real coefficients depending on the initial ($t=0$) condition.
Eq.~\ref{spectral_decomposition:eqn} shows that the approach to the equilibrium distribution is governed by
a \emph{relaxation time} $t_{rel} \propto \bar{\lambda}_1^{-1}$, $\bar{\lambda}_1$ being the smallest non-zero eigenvalue 
of $-\bar{W}$ (recall that $\bar{\lambda}_0=0$).
Therefore, the larger the {\em spectral gap} $\Delta=\bar{\lambda}_1-\bar{\lambda}_0=\bar{\lambda}_1$, the faster the 
system reaches equilibrium.
However, since in the Metropolis algorithm the time step is actually discrete 
(see appendix~\ref{Model_spectrum:sec}),
it is better to consider the set of the $W$ eigenvalues, given by $\lambda_i = e^{-\Delta t\,\bar{\lambda}_i}$,
with $\lambda_0 =1$.
Therefore, in the rest of the paper we shall refer to the spectral gap of the original transition matrix 
$W$, which is $\Delta = 1-\lambda_1$, instead of the corresponding time-continuum limit 
introduced above. 
As we shall show in a while, this choice will not change the conclusion of our spectral analysis.

How can we use the spectral gap concept in an annealing context? 
It is natural to think that, for a given schedule $T(t)$, 
it would be good to {\em maximize} the spectral gap for each given temperature $T(t)$. 
This is indeed what we will explore in the following. 
Generally speaking, finding explicitly the spectral gap $\Delta$ for a given $\bar{W}$ is an impossible task; 
in the present one-dimensional double-well context, however, it
turns out that the spectrum of $W(x'|x)$ is quite simple to obtain by, for instance, 
a straightforward discretization of the real axis. 
 
We present now the results obtained by applying this spectral analysis to the proposal moves we defined 
in Sec.~\ref{CA_intro:sec}.
We recall that they are: The Box move, see Eq.~\ref{P_transition_matrix_box:eqn}, the Gaussian move, 
see Eq.~\ref{P_transition_matrix_gau:eqn}, and the Lorentzian one, see Eq.~\ref{P_transition_matrix_lor:eqn}.
The main quantity under inspection is the \emph{spectral gap} 
$\Delta$ of the transition matrix $W(x'|x)$ which is a 
function of both the temperature $T$ and the proposal move range $\sigma$, $\Delta(T,\sigma)$.
We obtain this gap (as well as the higher eigenvalues) by diagonalizing an appropriate symmetrized
(see appendix~\ref{Model_spectrum:sec}) and discretized version of the 
transition operator $W(x'|x)$. 
\cite{footnote1}
%
\begin{figure}[ht]
\begin{center}
\includegraphics[width=8cm]{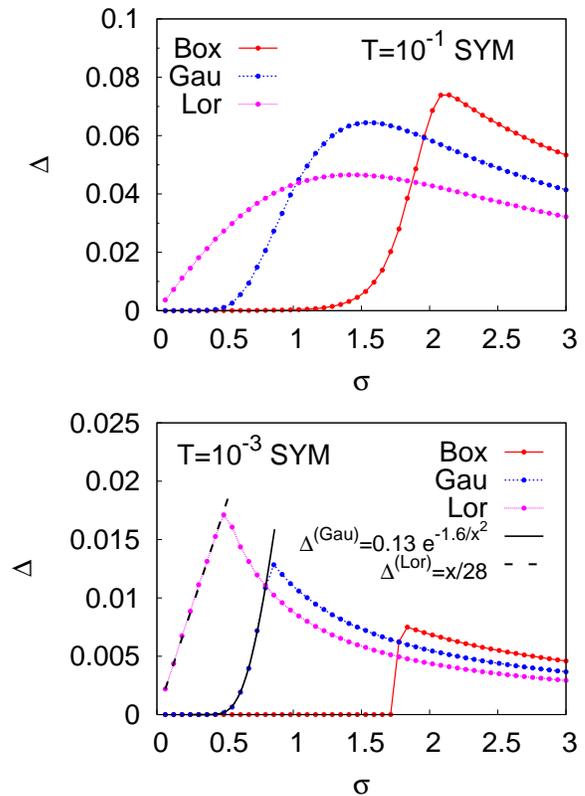}
\end{center}
\caption{
Exact diagonalization of the transition operator $W$, for the symmetric potential case. 
We plot the spectral gap $\Delta$ versus the proposal move range $\sigma$.
The temperature is $T/V_0=0.1$ (Top) or $T/V_0=0.001$ (Bottom), in the same units in which the barrier parameter 
is $V_0=1$.
The solid and dashed black lines represent the fit provided by Eq.~\protect\ref{theory_gap_gau:eqn} (Gaussian case) and 
Eq.~\ref{theory_gap_lor:eqn} (Lorentzian case). The functional form of those fits is recalled in the legend.}
\label{all_vs_sigma_tot:fig}
\end{figure}
%
In Fig.~\ref{all_vs_sigma_tot:fig} we plot the spectral gap $\Delta$ versus $\sigma$ at 
two fixed temperatures, $T/V_0=0.1$ (Top) and $T/V_0=0.001$ (Bottom). 
As a general remark, we note how the behavior of $\Delta(\sigma)$ is remarkably different
for different trial moves.
Interestingly, for every proposal move and every $T$, there is a value of $\sigma$ which maximizes the gap 
$\Delta$. 
Recalling that the relaxation time is proportional to $\Delta^{-1}$ we anticipate that, at every
temperature $T$, there is a $\sigma(T)$ which provides the fastest relaxation. 
Clearly this $\sigma(T)$ is \emph{a priori} a better choice than taking just 
$\sigma(T) = \sigma_0\,(T/T_0)^{1/2}$, as previously implemented in Sec.~\ref{CA_intro:sec}.
At low $T$ however the dependence of $\Delta$ on $\sigma$ is no longer smooth. 
In particular, for $T/V_0=0.001$, the Box trial move shows a sharp transition 
from vanishingly small gap values to 
finite ones, while both Gaussian and Lorentzian moves show a cusp maximum. 
At low temperatures, the maximum gap $\Delta^{(Lor)}_{max}$ of the Lorentzian move is larger than 
$\Delta^{(Gau)}_{max}$, which is in turn larger than $\Delta^{(Box)}_{max}$; correspondingly, for small 
$\sigma$, we have $\Delta^{(Lor)}(\sigma)>\Delta^{(Gau)}(\sigma)>\Delta^{(Box)}(\sigma)$.
This is, essentially, the reason for the faster relaxation dynamics of the Lorentzian move during annealing.
%
Notice also that the maximum value $\Delta_{max}$ of $\Delta$, decreases with temperature in all cases, 
as one can see by comparing the two panels in Fig.~\ref{all_vs_sigma_tot:fig}. 
Moreover the value of $\sigma$ that maximizes the Lorentzian and Gaussian gap decreases with temperature, 
while it seems to converge to a finite value (around $\sigma \approx 1.75$) for the Box case. 

\begin{figure}[ht]
\begin{center}
\includegraphics[width=8cm]{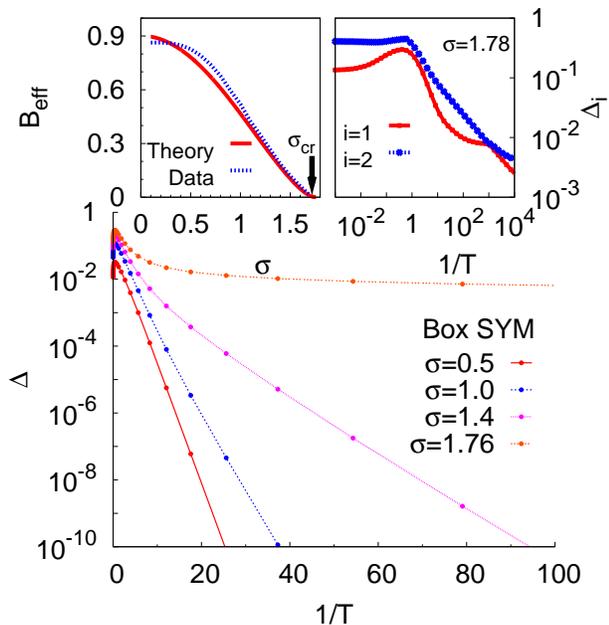}
\end{center}
\caption{
Exact diagonalization of the transition operator $W$, for the symmetric potential, 
with Box proposal move. We plot the spectral gap $\Delta$ versus the inverse temperature $1/T$ 
for several fixed values of the proposal move range $\sigma$.
Top left inset: The value of the effective barrier, $B_{\rm eff}$, seen by the system
when a Box proposal move with range $\sigma$ is employed. 
The value $\sigma_{cr}$ -- which corresponds to a vanishing effective barrier -- 
is indicated by a vertical arrow.
Top right inset: The two first gaps, $\Delta_1=1-\lambda_1$ and $\Delta_2=1-\lambda_2$, 
for a Box proposal move versus the inverse temperature, $1/T$, at fixed move range $\sigma=1.78$}.
\label{box_vs_T:fig}
\end{figure}
\begin{figure}[ht]
\begin{center}
\includegraphics[width=8cm]{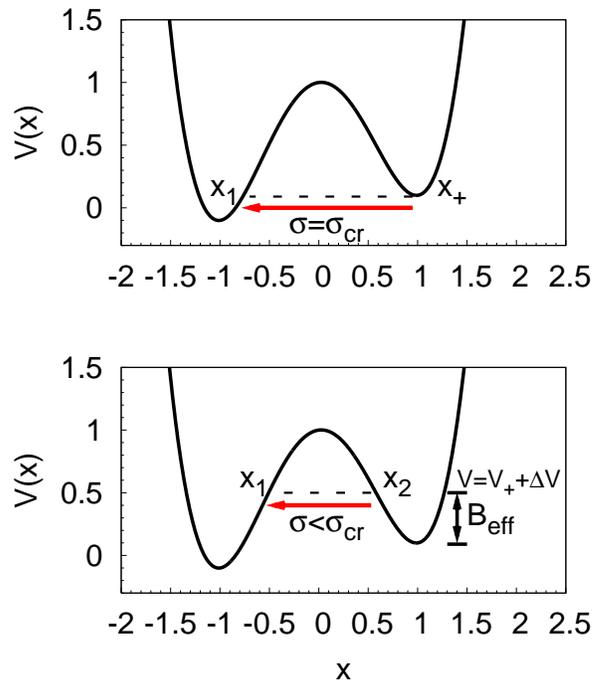}
\end{center}
\caption{
Illustrative sketch of a Box move in a double-well potential. In the upper panel we consider
the case $\sigma = \sigma_{cr}$, while in the lower one we illustrate the general case
$\sigma < \sigma_{cr}$. Here $V_+=V(x_+)$ is the potential at the bottom of the metastable
minimum. The meaning of the other symbols is explained in the text.
}
\label{fig_double_well_barrier:fig}
\end{figure}

%
The Box case is particularly intriguing, and deserves a closer inspection.
In Fig.~\ref{box_vs_T:fig} we plot $\Delta$ versus inverse temperature $1/T$ for the case
of the Box move, and for several values of $\sigma$. 
For a whole range of $\sigma$, $\sigma<\sigma_{cr}\approx 1.75$, we see a low-temperature behavior
typical of an activated Arrhenius process $\Delta(T) \propto e^{-B_{\rm eff}(\sigma)/T}$, 
where $B_{\rm eff}(\sigma)$ is an \emph{effective barrier}, clearly $\sigma$-dependent, which the system experiences. 
Above a certain $\sigma_{cr}$, the behavior of $\Delta$ changes drastically, from Arrhenius to what 
appears, at first sight, just a constant. A closer inspection on an extended temperature range, see
Fig.~\ref{box_vs_T:fig} (Top right inset), shows that at very small temperatures an (avoided) crossing between 
eigenvalues of $W$ has taken place, and $\Delta$ starts to decrease again toward zero 
as $\Delta\propto T^{1/2}$.
\cite{harmonic:note}
The effective barrier $B_{\rm eff}(\sigma)$ extracted from an Arrhenius fit, 
$\Delta \propto e^{-B_{\rm eff}(\sigma)/T}$, 
goes to zero for $\sigma\to \sigma_{cr}$.
There is a simple geometrical explanation for this critical behavior. 
First of all, we note that the value of $\sigma_{cr}$ is close to the distance between the two
minima. This suggests that the transition is related to the availability of a {\em direct jump} 
from the bottom of the metastable well to the other one.
If we call $x_{\pm}$, respectively, the minimum in the right and left well, there is a
non-trivial solution of the equation $V(x_1)=V(x_+)$ with $x_1$ lying between the
two minima, $x_-<x_1< x_+$ (see Fig.~\ref{fig_double_well_barrier:fig}): 
It turns out that $\sigma_{cr}=x_+-x_1$. 
%

Indeed, for $\sigma > \sigma_{cr}$ there is a possible proposed move that
brings the system from $x=x_+$ (the metastable minimum) to the point $x^{\prime}=x_1$ on the other
side of the barrier: Such a ``non-local'' move pays no energy ($\Delta E=0$), and is
therefore certainly accepted by the Metropolis algorithm.
In some sense, for $\sigma>\sigma_{cr}$ the barrier is not seen (or, at least, there are allowed
and accepted moves that do not see it), and $B_{\rm eff}$ is zero.
Consider now the case $\sigma<\sigma_{cr}$.
For every value $V=V(x_+)+\Delta V$, with $\Delta V>0$, there are two equipotential solutions 
$x_1(\Delta V)$ and $x_2(\Delta V)$, such that $V(x_1)=V(x_2)=V$, lying between the two minima, 
but on opposite sides of the barrier, i.e., with $x_-<x_1(\Delta V)<x_2(\Delta V)<x_+$.
Denote now by $d(\Delta V)=x_2(\Delta V)-x_1(\Delta V)$ the distance between such two 
equipotential points ($d(\Delta V)$ is a monotonically decreasing function of $\Delta V$).
If the box width is $\sigma$, we can always find $\Delta V$ such that $\sigma=d(\Delta V)$,
i.e., such that there is a proposed move connecting $x=x_2$ to $x^\prime=x_1$ which, once again,
bypasses the top of the barrier.
The effective barrier seen by the system close to the metastable minimum, for a given value of $\sigma$,
is therefore just (see Fig.~\ref{fig_double_well_barrier:fig}) 
\begin{equation} \label{B_eff_box:eqn}
B_{\rm eff}(\sigma) = \Delta V = d^{-1}(\sigma) \;,
\end{equation}
i.e., in essence, a piece of length $\sigma$ is cut from the top of the barrier, 
and is effectively not seen by the system.
In other words, the effective barrier $B_{\rm eff}$ is just the potential energy drop $\Delta V$ 
which the system must overcome before a long jump $|x'-x|\sim \sigma$ is made available at no energy cost
($V(x')\sim V(x)$).
The value of $B_{\rm eff}(\sigma)$ obtained through such a simple geometric construction is shown 
by a solid line in Fig.~\ref{box_vs_T:fig}(Top left inset): The agreement with the numerical data 
-- extracted from the Arrhenius fit --
is remarkably
good, with small deviations for very small values of $\sigma$, which are likely due to the effect of the finite grid employed in diagonalizing the transition operator $W$ (see Sec.~\ref{Spectral:subsec}),
and to finite $T$ effects.
\cite{footnote2}
%

We would like stress that this geometrical picture is strictly true only at zero temperature, 
and does not apply to the Gaussian or Lorentzian cases, whose tails provide a small 
non-vanishing chance of making a ``long jump'' for any value of $\sigma$. 
The treatment of the Gaussian and Lorentzian cases needs, therefore, a more specific 
and technical discussion, 
which we sketch in appendix \ref{Model_spectrum:sec}.

Summarizing, the Box trial move shows a {\em sharp} (first-order-like) transition at a value 
of $\sigma_{cr}\approx 1.75$, where the effective Arrhenius barrier 
$B_{\rm eff}(\sigma)$ vanishes, and the gap starts to decrease as a power-law, $\Delta\propto T^{1/2}$,
for very small values of $T$  (see Fig.~\ref{box_vs_T:fig}). 

One might suspect that the cusps shown in Fig.~\ref{all_vs_sigma_tot:fig} for the Gaussian and Lorentzian cases
signal, similarly to the Box case, some kind of transition. 
A closer inspection, however, shows that this is not so: It is just a level crossing phenomenon.
%
\begin{figure}[ht]
\begin{center}
\includegraphics[width=8cm]{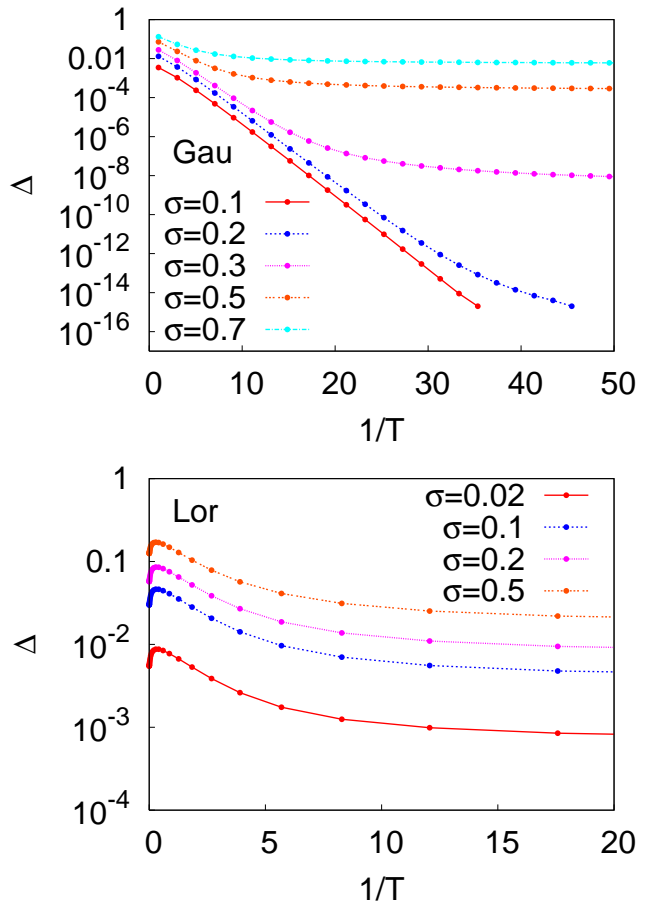}
\end{center}
\caption{
Exact diagonalization of the transition operator, for the symmetric potential, with Gaussian (Top)
or Lorentzian (Bottom) move. 
We plot the gap $\Delta$ versus the inverse temperature $1/T$ for several fixed values of the proposal move
range $\sigma$.}
\label{gau_plus_lor_vs_T:fig}
\end{figure}
%
In Fig.~\ref{gau_plus_lor_vs_T:fig} we plot $\Delta$ versus the inverse temperature $1/T$
for several values of $\sigma$, for the Gaussian (Top panel) and Lorentzian (Bottom panel) proposal moves.
After an initial Arrhenius-like behavior, particularly visible for the Gaussian small $\sigma$ cases, 
the system always, and smoothly, changes to a low-$T$ behavior which is entirely similar to the
large-$\sigma$ Box case: An apparent saturation of $\Delta$ followed by an avoided crossing
between the first two exited states (not shown) and a final $\Delta\propto T^{1/2}$. 
\cite{harmonic:note}
No real transition exist in the Gaussian and Lorentzian case: The cusp in Fig.~\ref{all_vs_sigma_tot:fig} 
moves down toward smaller and smaller values of $\sigma$ for decreasing $T$. 

\subsection{Classical Annealing with optimal $\sigma(T)$}
\label{Optimal_schedule:subsec}
%
In the previous section we argued about the possibility of an optimal choice for the proposal move range 
$\sigma(T)$. This choice should guarantee the fastest \emph{instantaneous} relaxation toward
the instantaneous equilibrium distribution $P_{eq}(x)$ at any given temperature. 
The classical annealing performance is expected to be greatly improved by such a choice of $\sigma(T)$.

In practice, for each choice of proposal move (Box, Gaussian, Lorentzian), we numerically
determined the value of $\sigma$ that maximizes the spectral gap $\Delta$ for each fixed temperature $T$.
We will refer here to such an optimal $\sigma$ as $\sigma_{opt}(T)$. 
%
\begin{figure}[ht]
\begin{center}
\includegraphics[width=8cm]{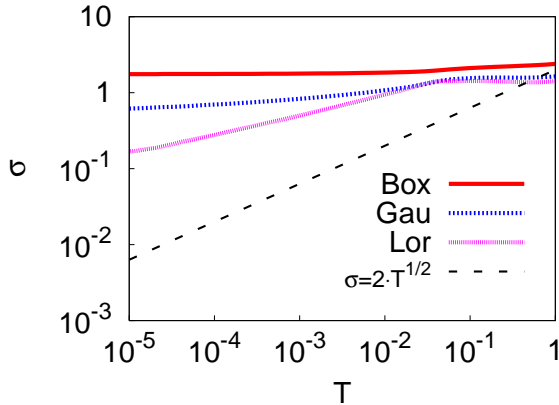}
\end{center}
\caption{
Plot of the optimal $\sigma_{opt}(T)$ for the Box, Gaussian and Lorentzian 
proposal move. We also show, for comparison, the schedule $\sigma(T) \propto T^{1/2}$ 
employed in Sec.~\ref{CA_intro:sec}.
}
\label{fig_optimal_sigma_vs_T:fig}
\end{figure}
%
In Fig.~\ref{fig_optimal_sigma_vs_T:fig} we plot such an optimal choice for all the
types of proposal moves introduced in Sec.~\ref{CA_intro:sec}. For comparison, we also show the
$\sigma(T)$ we employed in Sec.~\ref{CA_intro:sec}, which is definitely smaller than $\sigma_{opt}$, for
given value of $T$.

Having done this, we have then performed classical MC annealing runs, similarly to those reported in
Sec.~\ref{CA_intro:sec}, with the usual linear annealing schedule for the temperature 
$T(t)=T_0\left(1-t/\tau\right)$, where $\tau$ is the annealing time.
%
\begin{figure}[ht]
\begin{center}
\includegraphics[width=8cm]{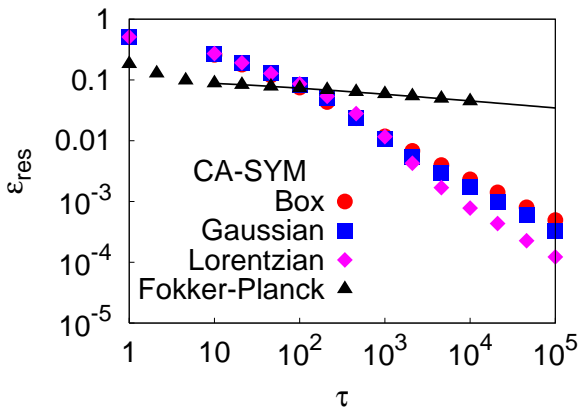}
\end{center}
\caption{
Plot of Monte Carlo classical annealings for the symmetric potential.
$\tau$ is the annealing time and $\epsilon_{res}$ the residual energy (see text). 
We report the results of the exact integration (Fokker-Planck), together with the actual MC data
for several proposal moves (Box, Gaussian, Lorentzian). The MC simulations are performed with
an optimal choice for $\sigma(T)$ obtained from the maximum instantaneous gap (see text).}
\label{FP_vs_MC_SYM_opt:fig}
\end{figure}
%
In Fig.~\ref{FP_vs_MC_SYM_opt:fig} we show the MC annealing results.
We stress again that the difference with respect to the runs illustrated in Fig.~\ref{FP_vs_MC_SYM_fdt:fig}, 
where we took $\sigma=\sigma_0(T/T_0)^{1/2}$, is that, here, $\sigma_{opt}(T)$ is the optimal choice of $\sigma$ 
obtained from the maximum instantaneous spectral gap of the transition matrix. 
We notice that the Box and Gaussian results are now very different form the previous ones
(see Fig.~\ref{FP_vs_MC_SYM_fdt:fig}).
Moreover, interestingly, Box and Gaussian data fall almost on the same curve, which is in turn very close
to the Lorentzian case, the latter showing once again the best annealing results.
\cite{footnote3}
\section{Quantum Annealing: Path-Integral Monte Carlo of the double-well} \label{PIMC:sec}
%
In I (Ref.~\onlinecite{Stella_simple:article}) we studied, by direct numerical integration, the 
real- and imaginary-time Schr\"odinger dynamics provided by the \emph{time-dependent} Hamiltonian 
\begin{equation} \label{quadratic_hamiltonian:eqn}
H = -\Gamma(t)\,\frac{\partial^2}{\partial \, x^2} + V(x) \;,
\end{equation}
where $\Gamma(t) = \hbar^2/[2m(t)] = \Gamma_0\,\left( 1-\frac{t}{\tau} \right)$ is the
inverse mass parameter appearing in the usual non-relativistic kinetic energy, providing the strength 
of the quantum fluctuations.
This is a literal implementation of the Quantum Annealing (QA) strategy.
However as pointed out in the introduction, this Schr\"odinger dynamics 
is suitable only for toy problems with a very limited Hilbert space, and quite inapplicable 
for actual optimization. 
In order to be a viable strategy for realistic optimization, QA must resort to {\em stochastic}, 
i.e., Quantum Monte Carlo (QMC), approaches, mostly appropriate for an {\em imaginary-time} framework 
(we remind the reader that, as shown in I, working in imaginary-time is actually beneficial for QA). 

There are several possible QMC techniques on which a QA strategy can be build. 
By far, the simplest of these QMC techniques is the {\em Path-Integral Monte Carlo} (PIMC) method, 
which has already been used with some success in the QA 
context \cite{Santoro:science,Martonak:ising,Martonak:TSP,demian_QA:article}.
The method does not addresses the imaginary-time Schr\"odinger time-dependent dynamics, but simulates an
{\em equilibrium quantum system}, held at a small finite temperature $T$, where the relevant 
quantum parameter $\Gamma(t)$ 
is externally switched off in the course of a QMC simulation: 
$t$ is not treated, therefore, as a proper physical time, and is only replaced by 
a {\em Monte Carlo time}.
\cite{footnote4}

In the present section we will explore the potential of a PIMC-based QA strategy on the same simple
one-dimensional double-well potential. The reason for investing so much effort on this
simple problem, is that the simplicity of the problem landscape will allow us to have 
a perfect control of all the ingredients of the method. Indeed, we will learn a lot 
about PIMC-QA, particularly about the limitations of the
method, from investigating this toy problem.

The equilibrium properties of our quantum system, for any fixed value of $\Gamma=\Gamma(t)$, are all encoded in 
the quantum partition function 
\begin{equation} \label{partitionfunction:eqn}
Z(\beta) = {\rm Tr} \left[e^{-\beta\,(\hat{T}+\hat{V})}\right] = 
\int dx \; \langle x| e^{-\beta(\hat{T}+\hat{V})} |x \rangle 
\end{equation}
where
\begin{equation} 
\hat{T} = -\Gamma\,\frac{\partial^2}{\partial \, x^2}  \;,
\end{equation}
from which all the thermodynamics follows. 
%
Here, as usual, $\beta=(k_B\,T)^{-1}$. We will drop the Boltzmann constant $k_B$ from now on.
As we see from Eq.~\ref{partitionfunction:eqn}, $Z$ involves an integral over a positive distribution,
$\langle x| e^{-\beta(\hat{T}+\hat{V})} |x \rangle$, which involves, however, the very difficult task 
of calculating the diagonal matrix element of the {\em exponential} of the Hamiltonian $H$. 

The standard approach is to rewrite Eq.~\ref{partitionfunction:eqn} as a Feynman 
\emph{Path-Integral} \cite{FeynmanSTAT:book}.
The main mathematical tool employed in such an approach is the so-called 
\emph{Trotter theorem}, which reads:
\begin{equation} \label{trotter_theorem:eqn}
e^{-\beta(\hat{T}+\hat{V})} = \lim_{P\to\infty} 
\left( e^{-\frac{\beta}{P}\,\hat{T}}\,e^{-\frac{\beta}{P}\,\hat{V}} \right)^{P} \;,
\end{equation}
where $P$ is the number of \emph{Trotter's slices (or replicas)}, in which the imaginary time interval 
$[0,\hbar\beta]$ has been partitioned, each slice being of length $\Delta\,t=(\hbar\,\beta)/P$.
By means of Eq.~\ref{trotter_theorem:eqn}, and inserting $P-1$ identities in the form of 
${\bf 1}_i=\int dx_i |x_i \rangle \langle x_i|$, one can rewrite the partition function
(see, for instance, Ref.~\onlinecite{CeperleyPIMC:review} for this simple derivation) 
as:
\begin{equation} \label{primitive_approx_text:eqn}
Z(\beta) = \left(\frac{1}{4\,\pi\,\Gamma\,\Delta\,t}\right)^{\frac{P}{2}}\,\int\,\prod_{i=0}^{P-1}\,{\rm d}x_i\,
e^{-S_{PA}[x]} + O\left(\Delta\,t^2\,\beta\right) \;,
\end{equation}
where $S_{PA}[x]$ is the so-called (euclidean) \emph{primitive action}: 
\begin{equation} \label{primitive_action_text:eqn}
S_{PA}[x] = \Delta\,t\,\sum_{i=0}^{P-1}\,
\left\{ \frac{m}{2}\,\left( \frac{x_{i+1}-x_i}{\Delta\,t} \right)^2 + V(x_i) \right\} \;,
\end{equation}
with periodic boundary conditions $x_{P}=x_0$, as
a consequence of the trace present in Eq.~\ref{partitionfunction:eqn}.)

The similarity of Eq.~\ref{primitive_approx_text:eqn} with the classical partition function of a 
{\em closed polymer} with $P$ beads is evident: The polymer beads $x_i$ can be seen as the
imaginary-time snapshots $x_i=x(i\Delta\,t)$, $i=0,\cdots,P-1$, of a fluctuating closed path $x(t)$ in 
the enlarged configuration space $(x,t)$, where $t$ is the imaginary time. 
Two neighboring beads $x_i$ and $x_{i+1}$ interact with harmonic interactions of spring constant 
$K^{\perp}=mP^2/(\hbar\beta)^2$, originating from the propagator of the quantum kinetic term, 
\begin{equation} \label{Hkin_primitive:eqn}
\langle x_i | e^{-\frac{\beta}{P}\,\hat{T}} | x_{i+1} \rangle \propto
e^{ -\frac{\beta}{P} \frac{K^{\perp}}{2} (x_{i+1}-x_i)^2 } \;.
\end{equation}
%
The strength of the harmonic interactions in imaginary-time controls the amount of quantum fluctuations:
A large mass $m$ (classical regime) results in a strong $K^{\perp}$ and hence in a very ``rigid'' polymer, 
while a small mass $m$ (quantum regime) results in a very soft and fluctuating polymer. 
All beads are also subject to the classical potential $V(x)$. 
Once we have reduced our problem to the Path-Integral form in Eq.~\ref{primitive_approx_text:eqn}, the
standard techniques of classical Metropolis MC can be used, and the resulting algorithm is
what is called a {\em Path-Integral Monte Carlo} (PIMC). 
The most obvious MC moves to be used are just {\em single bead} moves, exactly
as in a classical MC. 

These are the bare bones of a PIMC approach. 
For the problem we are dealing with, a particle in a potential (or, more generally, for systems of 
quantum particles on the continuum), one can improve on the method just sketched in two possible directions: 
{\em i)} by improving the quality of the approximation in Eq.~\ref{primitive_approx_text:eqn}, so as to get
a smaller Trotter discretization error \cite{Suzuki:book} for a given number of Trotter slices $P$ used; 
\cite{footnote5}
{\em ii)} by introducing MC moves which are more sophisticated than just moving a single bead at a time. 
Regarding {\em i)}, we will adopt a {\em fourth order} approximation to the 
action \cite{Takahashi_Imada:article,Li_Broughton:article}, which improves the Trotter
truncation error of the partition function
from $O\left(\Delta\,t^2\,\beta\right)$ to $O\left(\Delta\,t^4\,\beta\right)$. 
%
%
More precisely, we used the so-called \emph{Takahashi-Imada} approximation 
\cite{Takahashi_Imada:article,Li_Broughton:article}, which is especially suitable 
for continuous systems.
The partition function is still given by an expression of the form Eq.~\ref{primitive_approx_text:eqn} 
with the primitive action $S_{PA}$ replaced by the following Takahashi-Imada action:
\begin{equation} \label{TakahashiImada_approx_text:eqn}
S_{TIA}[x] = \Delta\,t\, \sum_{i=0}^{P-1}\, \left\{ \frac{m}{2}\,
\left( \frac{x_{i+1}-x_i}{\Delta\,t} \right)^2 + V_{\rm eff}(x_i) \right\} \;,
\end{equation}
where the only difference with respect to $S_{PA}$ is in the potential energy, which now reads:
\begin{equation} \label{TakahashiImada_potential_text:eqn}
V_{\rm eff}(x) = V(x) + {1 \over 12} \Gamma \left( \Delta\,t \right)^2 \,
\left( \frac{\partial V(x)}{\partial\,x} \right)^2 \;.
\end{equation}
As for the MC move choice, {\em ii)}, we adopted smarter moves, known in the context of classical 
polymer simulations, which reconstruct entire pieces of the polymer, instead of a single bead at a time,
through the {\em bisection algorithm} \cite{CeperleyPIMC:review,Brualla:article}. 
This choice guarantees a fast relaxation to the instantaneous equilibrium distribution. 
Moreover, we also used global classical moves, in the form of rigid motions of the center of mass
of the polymer \cite{CeperleyPIMC:review}, so as to keep a good sampling in the final part of the
annealing, where the mass $m$ is large ($\Gamma$ is small) and quantum moves are suppressed.
%
%
In the next Section, we will present the annealing results obtained with such an allegedly
{\em state-of-the-art} PIMC.

\subsection{PIMC QA Implementation I: IV-order action and bisection moves}\label{results1:subsec}

Quantum annealing is performed by decreasing 
the inverse-mass parameter, $\Gamma(t)= \hbar^2/[2\,m(t)]$, appearing in 
Eq.~\ref{quadratic_hamiltonian:eqn}, linearly to zero in a time $\tau$, 
$\Gamma(t)=\Gamma_{0}\,\left(1-t/\tau \right)$
(it is understood here that both $t$ and the total annealing time $\tau$ are MC times, 
i.e., measured in units of MC steps, each MC step consisting of one bisection move plus one 
global move). 
The initial condition is set to $\Gamma_{0}=0.5$ (as in I). 
For every value of the annealing time $\tau$ we calculated the relevant averages by repeating several 
times the same annealing experiment, starting from different randomly distributed initial conditions.
\cite{footnote6}

\begin{figure}[ht]
\begin{center}
\includegraphics[width=8cm]{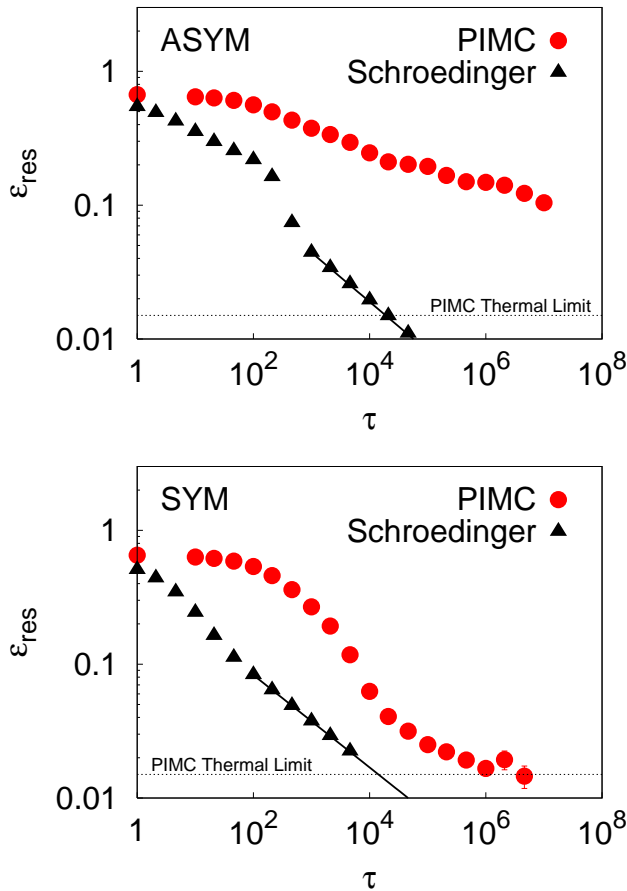}
\end{center}
\caption{
(Top) PIMC-QA residual energy for the asymmetric potential, using a fourth-order action and the 
bisection algorithm. 
The dashed line indicates the thermal equipartition limit $T/2$, for $T/V_0=0.03$.
As a reference, the results obtained in Ref.~\onlinecite{Stella_simple:article} by exact integration
of the imaginary-time Schr\"odinger equation are reported. %
(Bottom) Same as above, for a symmetric potential.} 
\label{SE_vs_MC_nolor_noins:fig}
\end{figure}

In Fig.~\ref{SE_vs_MC_nolor_noins:fig} we plot the final PIMC-QA residual energy obtained for both potential 
choices, asymmetric (Top) and symmetric (Bottom) 
(with the same parameters used in I and in Sec.~\ref{CA_intro:sec}), 
for a fixed temperature $T/V_0=0.03$.
The statistical errors are evaluated by $10^3$ repetitions of each annealing run for $\tau<10^5$, and $10^2$ 
repetitions for $\tau>10^5$.
We employed $P=160$ with an $l=5$ bisection level (i.e., moving polymer pieces containing $2^5-1=31$ beads
\cite{CeperleyPIMC:review}, for $\tau\leq10^5$, and $P=20$ with $l=2$ (moving $2^2-1=3$ beads) for $\tau>10^5$. 
We stress that the two series of data match perfectly, but obviously the use of a smaller $P$ allows us to achieve
larger annealing times $\tau$, which would be otherwise too computationally heavy.
(An analysis of the convergence behavior of the algorithm as a function of both $\Gamma$ and $P$ can be found 
in Ref.~\onlinecite{Stella:phdthesis}.)
%
%
In the asymmetric potential case (Top panel of Fig.~\ref{SE_vs_MC_nolor_noins:fig})
-- which, we recall, presents a level crossing and a Landau-Zener (LZ) transition (see I) --, 
the residual energy $\epsilon_{res}(\tau)$ as a function of the annealing time $\tau$ 
gets stuck around $\epsilon_{res}=0.2$. 
Since this energy is comparable to the energy of the metastable state, 
$V(x_{+})$, and definitely larger than the
thermal limit $T/2=0.015 V_0$, we see that the algorithm failed to follow adiabatically the ground state.
For comparison, we also reported in Fig.~\ref{SE_vs_MC_nolor_noins:fig} the residual energy data obtained
by the exact imaginary-time Schr\"odinger annealing (reported in I), 
even if the time-scales of the two algorithms are definitely different and unrelated.
Considering, on the other hand, the symmetric potential case, bottom panel of 
Fig.~\ref{SE_vs_MC_nolor_noins:fig}, we notice that the algorithm succeeds in reaching the thermal limit 
$T/2$ in a reasonable amount of MC steps. 

From this first comparison between PIMC-QA and exact Schr\"odinger QA, we appreciate that the LZ
transition leads to a {\em severe slowing down} of the Monte Carlo algorithm, the actual tunneling event being
essentially missed by the PIMC algorithm. 
This difficulty is also present in \emph{static} simulations performed in the neighborhood of the LZ crossing 
(occurring at $\Gamma \sim 0.038$), where we find (results not shown) a dangerous loss
of adiabaticity which calls for an improved sampling of the action: We will show how this improved sampling
is achieved by the introduction of {\em instanton moves}. 

\subsection{PIMC Implementation II: Adding the Instanton move}\label{results2:subsec}
In the previous section we tested an allegedly state-of-the-art PIMC-QA algorithm for the very simple problem
of a particle in a double-well potential, with disappointing results. 
The problem is a {\em sampling crisis}: Our action is accurate, but its sampling misses
the tunneling events, which is catastrophic when a Landau-Zener crossing occurs 
(asymmetric potential case).
The well-known cure for this kind of sampling problem, for the case of a \emph{perfectly symmetric} double-well,
$V(x) = V_0 (x^2-a^2)$,
is the introduction of the so-called \emph{instanton move} (see Ref.~\onlinecite{Negele_Orland:book}).
In a nutshell, an instanton is defined as a solution of the classical equation of 
motion in the inverted potential, 
which goes from one minimum to the other. Moreover, the time-reversed path (anti-instanton) is also a solution. 
Instanton solutions are given by:
\begin{equation} \label{instanton_def:eqn}
x_{cl}(\tau)= \pm\,\tanh{\left( \frac{\omega_{ins}\,(\tau-\tau_0)}{2}\right)}\;
\end{equation}
where the $\pm$ signs denote instanton and anti-instanton, respectively, $\omega_{ins} = (8\,\Gamma\,V_0)/a^2$
is the instanton frequency, and $\tau_0$ a configuration parameter (the instanton center).
Strictly speaking, the whole classical trajectory takes an infinite (real) time, 
while the barrier overcoming is 
a very fast process (whence the name instanton). 
In our implementation, we made use of the imaginary-time version of an instanton/anti-instanton pair as 
a proposal MC move \cite{Stella:phdthesis,Negele_Orland:book}. 
In practice, the instanton move proposes to a subset 
of the Trotter's slices an excursion from one minimum to the other. The move will be accepted according
to the usual Metropolis criterion, with an energy competition between the possible 
gain in potential energy compensated
by the increase in kinetic energy due to the spring stretchings (at the positions of the instanton and of the 
anti-instanton).
Obviously, the instanton move will be not effective in the classical limit (small $\Gamma$, final part of the
annealing), since it describes an inherently quantum effect: The tunneling process between the two wells.
In particular, the instanton frequency, $\omega_{ins}$ (defined in Eq.~\ref{instanton_def:eqn}), decreases with $\Gamma$,
and when $1/\omega_{ins} \gtrsim \beta/2$, the tunneling time will exceed the thermal cut-off.
As a consequence, the instanton/anti-instanton pair will reduce to a useless flat configuration 
over the interval $[0,\hbar\,\beta]$.

Before showing the results, we need to stress an important point: The instanton move is available 
only for potentials which are small deformations of a perfectly symmetric double-well potential 
-- as our symmetric and asymmetric cases are, see Sec.~\ref{CA_intro:sec} -- and is not 
the general key for solving 
sampling problems (ergodicity breaking) for a generic potential, whose landscape is generally poorly known. 
In essence, we are playing here, for demonstration purposes, an unfair game, 
using a {\em detailed information} on
the potential landscape in order to correctly implement the tunneling dynamics in our PIMC. 

\begin{figure}[ht]
\begin{center}
\includegraphics[width=8cm]{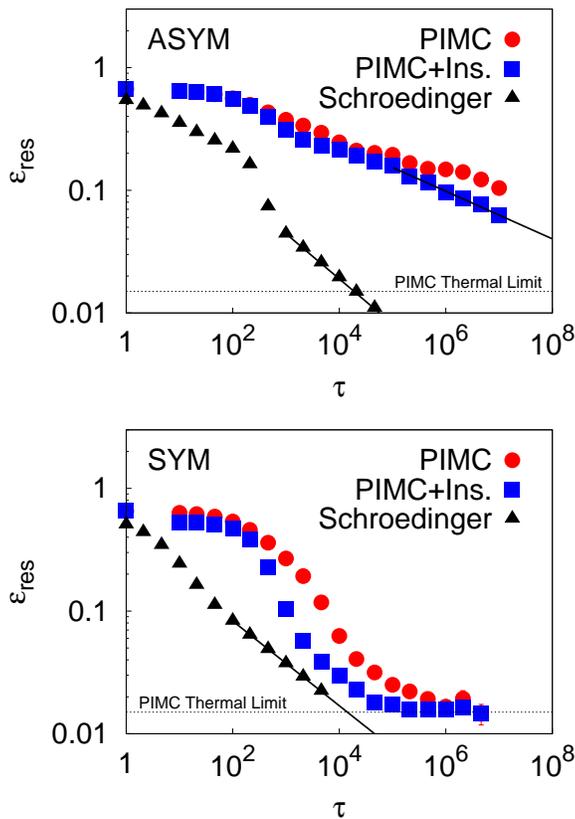}
\end{center}
\caption{
(Top and Bottom) Same as in Fig.~\protect\ref{SE_vs_MC_nolor_noins:fig}, with the instanton move allowed.
As a reference the results obtained without instanton move are still reported.
The data fit for the asymmetric potential case is discussed in the text.}
\label{SE_vs_MC_nolor:fig}
\end{figure}

Fig.~\ref{SE_vs_MC_nolor:fig} shows the PIMC-QA residual energy results when the instanton move is introduced, 
for the case of asymmetric (Top) and symmetric potential (Bottom). 
As in the previous section, we employed $P=160$ Trotter's slices with $l=5$ bisection level, for $\tau\leq10^5$, 
and $P=20$ with $l=2$ bisection level for $\tau>10^5$. 
Statistical errors are evaluated with $10^3$ repetitions of every annealing run. 
%
%

We note how the introduction of the instanton move causes a visible 
improvement of the residual energy slope for the asymmetric potential case:  
The asymptotic power-law behavior is now quite evident, $\epsilon_{res}(\tau)\propto \tau^{-\Omega_{QA}}$,
although with an exponent which is only $\Omega_{QA}=0.19$ (smaller than in the Schr\"odinger integration case,
where $\Omega_{QA}=1/3$, see I).
For the symmetric potential case, the instanton move leads to a faster convergence to the 
thermal limit, but does not really give an overwhelming  qualitative change. 
In particular we note that the partial acceptance ratio of the instanton moves is very low -- around $1\%$ --, 
despite the fact that this proposal move was  ``tailored'' on the potential under investigation.
Nevertheless, the instanton move quantitatively changes the PIMC-QA performance. 
This is perhaps catch a generic feature of systems with barriers, namely the 
crucial importance of rare events -- tunneling in this case.

Unfortunately, as said, the instanton ``recipe'', natural as it is for a 
double-well potential, cannot be generalized
to generic landscapes, including more complicated potentials, not to mention actual combinatorial optimization problem.
Nevertheless, the results obtained are instructive in many respects: They show 
clearly an important limitation
of the PIMC approach, while demonstrating, once more, that a smart choice of the proposal move --
even in the case of PIMC-QA -- leads to immediate improvement of the annealing performance.  

\subsection{PIMC Implementation III: Choosing the right kinetic energy. The Lorentzian move.}\label{results3:subsec}
%
It is by now a sort of {\em Leitmotiv} of the paper, that the choice
of the MC move strongly influences the dynamics, and hence the annealing behavior.
We recall, in particular, see Sec.~\ref{CA_intro:sec}, that in the classical annealing
case the winning choice was given by a Lorentzian distributed proposal move, which 
sometimes provides very long displacements. 

In the present PIMC context, however, the non-relativistic form of the kinetic energy 
part of the Hamiltonian
Eq.~\ref{quadratic_hamiltonian:eqn} {\em forces}, in a sense, the choice of Gaussian distributed moves, 
because the free propagator of the non-relativistic kinetic energy is a Gaussian, 
see Eq.~\ref{Hkin_primitive:eqn}.
The whole bisection algorithm makes strong use \cite{CeperleyPIMC:review}
of the Gaussian nature of the free propagator.

It is natural to ask what would be the QA behavior if, instead of the standard 
non-relativistic kinetic energy used so far, we employ for example 
a \emph{relativistic} Hamiltonian of the form
\begin{equation} \label{relativistic_hamiltonian:eqn}
H(t) = \Gamma(t)\,|p| + V(x) \;.
\end{equation}
One immediate consequence of this choice is that the free Gaussian propagator appearing in
Eq.~\ref{Hkin_primitive:eqn} is now replaced by a Lorentzian:
\begin{equation} \label{Hkin_lorentzian:eqn}
\langle x_i | e^{-\Delta\,t\,\Gamma |p|} | x_{i+1} \rangle =
\frac{1}{\pi}\,
\frac{\Delta\,t\,\Gamma}{(\Delta\,t\,\Gamma)^2 + (x_i-x_{i+1})^2 } \;.
\end{equation} 
With a certain effort, we have succeeded in generalizing the bisection algorithm to implement
in a very effective way the dynamics provided by this Lorentzian propagator 
(see appendix \ref{lorentzian_move:sec} and Ref.~\onlinecite{Stella:phdthesis} for more details).

\begin{figure}[ht]
\begin{center}
\includegraphics[width=8cm]{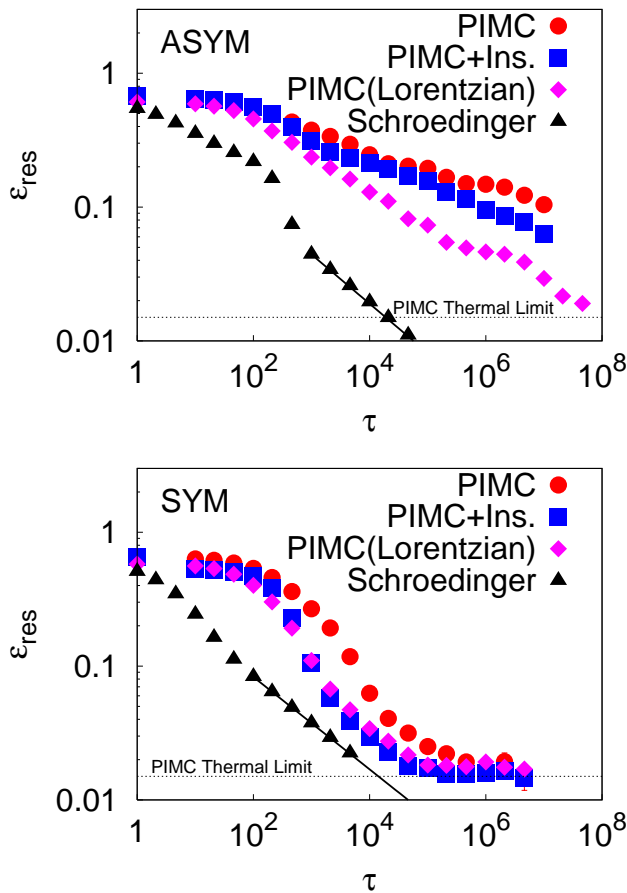}
\end{center}
\caption{
Same as in Fig.~\protect\ref{SE_vs_MC_nolor_noins:fig}, but for a QA based on the relativistic kinetic
energy of Eq.~\protect\ref{relativistic_hamiltonian:eqn}, implemented via a bisection algorithm adapted 
to Lorentzian moves.
As a reference, the results obtained by exact integration of the imaginary-time Schr\"odinger and by the 
previous Gaussian-based PIMC-QA with and without instanton move are still reported.} 
\label{SE_vs_MC:fig}
\end{figure}

We present here the results of a PIMC-QA approach which implements 
the relativistic kinetic energy, through a bisection algorithm with such a ``Lorentzian move''. 
In Fig.~\ref{SE_vs_MC:fig} we report the residual energy results for the case of the asymmetric
(Top) and symmetric potential (Bottom).
The results are obtained using $P=40$ Trotter's slices and $l=2$ bisection steps.
Averages and statistical errors are calculated, as usual, with $10^3$ repetitions of every annealing run.
It is clearly seen that, as in the classical case, the Lorentzian move -- 
{\it alias}, in the present context, the
{\em relativistic kinetic energy} -- greatly accelerates the QA behavior 
for the difficult asymmetric potential case. 
As usual, the ``simpler'' symmetric potential case does not show a qualitative change.
%

\section{Summary and discussion} \label{discussion:sec}
%
Summarizing, we would like to briefly stress some of the major conclusions reached in this paper.
\begin{description}
\item[1. {\it Role of move choice in Classical Annealing.}] The choice of the
Monte Carlo (MC) moves strongly influences the spectral properties of the 
MC Markov transition matrix, thus modifying
in a strong way the annealing properties. In some sense, the {\em effective landscape} of the
problem is really determined not only by the actual potential landscape, but also by the choses
Markov transition matrix, the latter determining the {\em neighborhood} of a given configuration 
\cite{theory_landscape:proceeding}.
In our double-well example a Metropolis MC with Lorentzian-distributed proposed moves ended up
showing the best annealing performances, overperforming not only the other classical MC choices, but
also any Path-Integral Monte Carlo (PIMC) algorithm which we were able to implement.

\item[2. {\it Tunneling and sampling difficulties of PIMC.}] A tunneling event, and its associated 
Landau-Zener crossing, can cause severe difficulties to a ``state-of-the-art'' PIMC-QA 
algorithm. The difficulty is associated to a poor sampling of the action 
which misses the rare, but all-important, tunneling events.
This problem can be cured by the ad-hoc introduction of ``instanton moves'', 
but this cure, while instructive, uses detailed information on the potential landscape 
that is generally not available.
The generalization of these ``instanton moves'' to more complicated potentials, let alone to generic hard
optimization problem, seems basically impossible, for it would require, among other things, 
a knowledge of the location of the minima were tunneling occurs.\\
 
\item[3. {\it Other limitations of PIMC.}] PIMC-QA suffers from a certain number of other limitations.
First of all, it works with a {\em finite temperature} $T$, and this sets up a thermal energy lower bound below
which we cannot possibly go (this limit was particularly clear in our double-well example). 
Second, a large number of Trotter slices $P$ can cause additional sampling problems in that an effective
uncorrelation of the configurations becomes harder and harder, even if a multi-step bisection algorithm is employed. 
Moreover, the {\em Trotter break-up} itself, see Eq.~\ref{trotter_theorem:eqn}, can cause difficulties
whenever, for a generic kinetic energy $T$, the form of the free propagator, generalizing
the simple Gaussian result of Eq.~\ref{Hkin_primitive:eqn}, is not known analytically.
This was indeed the difficulty met in a PIMC-QA study of the Traveling Salesman Problem  
\cite{Martonak:TSP,QA_Kolkata:proceeding}. 
 
\item[4. {\it Role of kinetic energy in QA.}] The choice of the kinetic energy is clearly all
important in QA: Sec.~\ref{results3:subsec}, illustrating the improvements in PIMC-QA 
upon using a relativistic kinetic energy, is particularly instructive. In our one-dimensional
example, this choice corresponds to a free propagator of Lorentzian form, which we implemented
by an appropriately modified bisection scheme (see App.~\ref{lorentzian_move:sec}).
%
%
The convenience of such a scheme in more general instances is not \emph{a priori} obvious and further tests on 
higher-dimensional problems would be needed.
\end{description}

In view of the previous points, it is fair to say that exploring alternative Quantum Monte Carlo schemes 
for performing stochastic implementations of Quantum Annealing remains an important and timely issue 
for future studies in this field. 
One such scheme, which in principle does not suffer from many of the
limitations of PIMC, is the Green's Function Monte Carlo scheme. Test applications of a GFMC-QA strategy
to the Ising spin glass are currently under way and will be reported shortly \cite{Stella_gfmc:article}.
%
%
\begin{acknowledgments}
This project was sponsored by MIUR through FIRB RBAU017S8R004, FIRB RBAU01LX5H,
PRIN/COFIN2003 and 2004, and by INFM (``Iniziativa trasversale calcolo parallelo'').
We acknowledge illuminating discussions with Demian Battaglia.
One of the author (LS) is grateful to Llorenc Brulla for his generous help in
discussing details about the PIMC algorithm.
\end{acknowledgments}
%
\appendix

\section{A model for the low-lying spectrum of the Metropolis transition operator}
\label{Model_spectrum:sec}
%
In this appendix we shall use analytical tools to explain some of the most relevant features of the 
Markov transition matrix spectra presented in Sec.~\ref{Spectral:subsec}. 
The question we want to address has to do with the low-lying spectrum of a Master equation of the form
\begin{equation} \label{master_eq:3_bis}
P_{n+1}(x') = \sum_x\, W(x'|x)\,P_{n}(x) \;,
\end{equation}
where $W(x'|x)$ is the transition operator associated to a certain choice of trial move in a Metropolis
MC scheme. 

As we have seen previously (see Sec.~\ref{Spectral:subsec}), the Box proposed move case admits 
a simple geometrical interpretation. 
For the general case, which is much harder to treat, we shall sketch here the main steps of a more systematic
procedure used to extract information on the gap of $W$, referring the reader to Ref.~\onlinecite{Stella:phdthesis}
for the many technical details. 

As explained in Sec.~\ref{Spectral:subsec}, 
it is possible to find a complete set of eigenvectors and eigenvalues of the
matrix $\bar{W}(i|j)$, defined through the short-time expansion of
this equation: $W=e^{\Delta t\,\bar{W}}$.
\cite{footnote7}
In particular, one can find a linear transformation from the (generally)
non-symmetric operator $-\bar{W}$ to a symmetric one, $\bar{H}$, 
which preserves its spectral properties \cite{VanKampen:book}, and
is easier to diagonalize in a standard way.
Moreover, one can apply the same linear transformation to the original transition matrix,
$W$; this leads to the
expansion of the master equation solution reported in Eq.~\ref{spectral_decomposition:eqn},
having taken the appropriate time-continuum limit.
However, in order to find the $W$ eigenvectors and eigenvalues
(and in particular its spectral gap, $\Delta$, defined in Sec.~\ref{Spectral:subsec}), 
it is better to keep $\Delta t$ finite and consider its simmetrized version,
that turns out to be the exponential operator 
$e^{-\Delta\,t\,\bar{H}}$,
where the appropriate $\Delta\,t$ is uniquely determined by the move range $\sigma$.
We notice that the spectrum of $-\bar{W}$, $\{ \bar{\lambda}_i\}$, introduced in Sec.~\ref{Spectral:subsec}
is simply related to the spectrum of the original transition matrix $W$, $\{ \lambda_i\}$, 
by the equation $\lambda_i=e^{-\Delta\,t\,\bar{\lambda}_i}$. 
In the following, we shall refer only to the gap of $W$. 

The potential $V(x)$ will be always assumed to have the usual double-well form. 
At very small temperatures, the equilibrium distribution $P_{eq}(x)$ is concentrated around the local minima at
$x=x_{\pm}$ \cite{Stella_simple:article}. 
Therefore, there is definitely a regime of parameters where a two-level system approximation for 
$W$ must hold. 
By exploiting this approximation, one can show (see Ref.~\onlinecite{Stella:phdthesis}) that the
spectral gap $\Delta=1-\lambda_1$ of the transition operator $W$ can be expressed, at small temperature $T$, 
in terms of a well-to-well propagator of the evolution operator of $\bar{H}$ as follows:
\begin{equation} \label{gap_value_integral:eqn}
\Delta \simeq \langle x_{-} |e^{-\Delta\,t\,\bar{H}}| x_{+} \rangle
\,e^{-\frac{V_{-}-V_{+}}{2\,T}} \;,
\end{equation}
where $V_{\pm}=V(x_{\pm})$. 
The crucial quantity we need, therefore, is the 
well-to-well propagator $\langle x_{-} |e^{-\Delta\,t\,\bar{H}}| x_{+} \rangle$. 
An expression for this propagator can be obtained by performing a Trotter decomposition and writing a Feynman 
path-integral \cite{FeynmanSTAT:book}.
The final result can be cast -- by an appropriate saddle-point approximation -- in the form:
\begin{widetext}
\begin{equation} \label{transition_amplitude_final:eqn}
\langle x_- |e^{-\Delta\,t\,\bar{H}}| x_+ \rangle \propto 
\int_{V_+}^{V_0}\,{\rm d}\,V \,\frac{\partial\,d}{\partial\,V}
\, K_1(d(V-V_+);\Delta\,t) \, e^{-\frac{1}{T}\,\left(V-\frac{V_+ + V_-}{2} \right)} \, P(d(V-V_+)) \;,
\end{equation}
\end{widetext}
where $d(V-V_+)$ is the function introduced in Sec.~\ref{Spectral:subsec} (see Fig.~\ref{fig_double_well_barrier:fig}),
and $K_1$ accounts for the Gaussian fluctuations around the saddle-point classical solution of the Feynman integral, 
which do not modify the important exponentially activated behavior provided by the term 
$e^{-\frac{1}{T}\,\left(V-\frac{V_+ + V_-}{2}\right)}$.
$P(d(V-V_+))$, finally, is just the proposal move distribution: $P(d(V-V_+))=P(|x_1-x_2|)=P(x_1|x_2)$
(see Fig.~\ref{fig_double_well_barrier:fig}).
 
This rather simple equation for the well-to-well propagator can be further treated by another standard saddle-point 
approximation of the relevant one-dimensional integral. We shall discuss here the cases we are interested in: 
Box, Gaussian, and Lorentzian. 

{\bf Box move}: $P$ was defined in Eq.~\ref{P_transition_matrix_box:eqn}. 
We recall that the function $d(V-V_+)$, introduced in Sec.~\ref{Spectral:subsec}, 
is a monotonically decreasing function of its argument, which exhibits an infinite first order derivative for $V=V_0$.
There is no true saddle-point for such a move, since the functions involved in
the exponential part of Eq.~\ref{transition_amplitude_final:eqn} are all monotonic. 
On the other hand, -- in the case of the Box move -- the extremes of integration in 
Eq.~\ref{transition_amplitude_final:eqn}
can be taken  as $[V_+,d^{-1}(\sigma)+V_+]$ instead of $[V_+,V_0]$,
since its transition matrix, $P(d(V-V_+))$, is zero outside the former interval.
Therefore, the largest contribution to the whole integral is due either to $V=V_0$ or $V=d^{-1}(\sigma)+V_+$,
which are the two integration extremes. 
It turns out that the latter choice guarantees the largest exponential, and is therefore the right one.
As a consequence, we can write down that:
\begin{equation}
\langle x_- |e^{-\Delta\,t\,\bar{H}}| x_+ \rangle \propto 
e^{-\frac{1}{T}\,\left(d^{-1}(\sigma) + \frac{V_+ - V_-}{2} \right)} \;.
\end{equation}
This transition amplitude is Arrhenius-like, and so does the gap function (see Eq.~\ref{gap_value_integral:eqn})
which reads:
\begin{equation}
\Delta(T,\sigma) \propto e^{-\frac{B_{\rm eff}(\sigma)}{T}} \;,
\end{equation}
where $B_{\rm eff}(\sigma)=d^{-1}(\sigma)$, as anticipated in Sec.~\ref{Spectral:subsec} from geometrical considerations. 
This is indeed the behavior seen in Fig.~\ref{box_vs_T:fig}.

{\bf Gaussian move}: $P$ was defined in Eq.~\ref{P_transition_matrix_gau:eqn}.
In this case, a true saddle-point is possible if the following equation has a solution:
\begin{equation} \label{saddle_point_gau:eqn}
\frac{1}{T} - \frac{d(V-V_+)}{\sigma^2}\,\frac{\partial\,}{\partial\,V}\,d(V-V_+) = 0 \;.
\end{equation}
Since $\frac{\partial\,d}{\partial\,V}<0$ and $d>0$, then, in the limit $T\to 0$ we have that
either $V=V_0$ or $V=V_+$ are solutions of Eq.~\ref{saddle_point_gau:eqn}.
The second choice guarantees the largest contribution to the integral in Eq.~\ref{transition_amplitude_final:eqn},
and must be therefore selected. The resulting expression for the transition amplitude is:
\begin{equation}
\langle x_- |e^{-\Delta\,t\,\bar{H}}| x_+ \rangle \propto e^{-\frac{V_+-V_-}{2\,T}}\,e^{-\frac{d(0)^2}{2\,\sigma^2}} \;,
\end{equation}
and the corresponding gap value is:
\begin{equation} \label{theory_gap_gau:eqn}
\Delta(T,\sigma) \propto e^{-\frac{d(0)^2}{2\,\sigma^2}} \;.
\end{equation}
Fig.~\ref{all_vs_sigma_tot:fig} shows that such a dependence perfectly fits the simulation data.
We emphasize that even the coefficient within the exponential part of the fitting function matches the theory 
(we recall that $d(0)=\sigma_{cr}\approx 1.75$): Only an overall prefactor is used as fitting parameter.

As observed in Sec.~\ref{Spectral:subsec}, the previous equation will be no longer valid for 
very small values of the temperature $T$, where an (avoided) crossing between the first eigenvalue $\lambda_1<1$
and the second eigenvalue $\lambda_2$ will occur.
On the other hand, for large values of $\sigma$ we see from Fig.~\ref{all_vs_sigma_tot:fig}
that the data deviate from the behavior predicted by Eq.~\ref{theory_gap_gau:eqn}. 
This is the consequence of another level crossing. 

{\bf Lorentzian move}: $P$ was defined in Eq.~\ref{P_transition_matrix_lor:eqn}. 
To cut a long story short (calculations proceed similarly to the Box and Gaussian case), we write down the result. 
The gap value is:
\begin{equation}\label{theory_gap_lor:eqn}
\Delta(T,\sigma) \propto \sigma \;,
\end{equation}
and also in this case the model agrees with the data, see Fig.~\ref{all_vs_sigma_tot:fig}, within a multiplicative 
fitting coefficient.

\section{The bisection algorithm with a Lorentzian Move}\label{lorentzian_move:sec}
%
In this appendix we want to show how to approach, within a Path-Integral Monte Carlo, 
a Hamiltonian with \emph{relativistic} kinetic energy:
\begin{equation} \label{Hrel:eqn}
H = \Gamma(t)\,|p| + V(x)\;.
\end{equation}
The kinetic operator $\Gamma(t)\,|p|$ is singular in a real space representation, 
but the corresponding density matrix operator can be written in a simple closed form:
\begin{equation}
\rho_K\left( x',x;\Delta\,t \right) = \frac{1}{\pi}\,\frac{\Gamma\,\Delta\,t}{\Gamma^2\,\Delta\,t^2+\left( x'-x\right)^2}\;.
\end{equation}
This is a \emph{Lorentzian (or Cauchy) distribution}, while the kinetic density matrix operator 
considered in Sec.~\ref{PIMC:sec}, Eq.~\ref{Hkin_primitive:eqn}, was a Gaussian one.

It is possible to obtain a generalization of the L\'evy construction
\cite{CeperleyPIMC:review} for this kind of Lorentzian move, observing that 
both Gaussian and Lorentzian distribution are stable under convolution (a property shared by all the so-called 
\emph{L\'evy distribution} \cite{Metzler:review}). 
The main idea is to obtain a closed form of the constrained kinetic operator:
\begin{equation}
T^{(1)}_{K}(x'_i) = \frac{ \rho_K\left( x_{i+1},x'_i;\Delta\,t \right)\,\rho_K\left( x'_i,x_{i-1};\Delta\,t \right)}
{\rho_K\left( x_{i+1},x_{i-1};2\,\Delta\,t \right)} \;,
\end{equation}
which is the probability of moving the $i$-th bead to $x'_i$, keeping fixed
the $(i+1)$-th and the $(i-1)$-th beads in $x_{i+1}$ and $x_{i-1}$, respectively.
We recall that this object is the building block of the bisection algorithm \cite{CeperleyPIMC:review}.
In the Lorentzian case, it reads:
\begin{equation}
T^{(1)}_{K}(x'_i) = \frac{\Gamma\,\Delta\,t}{2\,\pi}\,\frac{\left(\frac{x_{i+1}-x_{i-1}}{\Gamma\,\Delta\,t}\right)^2+4}
{\left[\left(\frac{x_{i+1}-x'_i}{\Gamma\,\Delta\,t}\right)^2+1\right]\,
\left[\left(\frac{x'_i-x_{i-1}}{\Gamma\,\Delta\,t}\right)^2+1\right]}\;.
\end{equation}
Unfortunately, this is no longer a Lorentzian distribution, and therefore it is not trivial to sample it 
(for instance, by making use of the usual transformation technique; see Ref.~\onlinecite{Kalos:book}).
However, after some algebra, and a smart change of variables (taking due care of the Jacobian of such a transformation,
which changes $T$ onto $\bar{T}$), we obtain a simpler form: 
\begin{equation}
\bar{T}^{(1)}_{K}(y) = \frac{2}{\pi}\,\frac{a^2+1}
{\left[\left(y+a\right)^2+1\right]\,\left[\left(y-a\right)^2+1\right]}\;,
\end{equation}
where $a = (x_{i+1}-x_{i-1})/(2\,\Gamma\,\Delta\,t)$, $c = (x_{i+1}+x_{i-1})/(2\,\Gamma\,\Delta\,t)$, and
\begin{equation} \label{change:eqn}
y = \frac{x'_i}{\Gamma\,\Delta\,t}-c \;. 
\end{equation}
Finally, with a bit of more algebra, we find that:
\begin{widetext}
\begin{equation}
\bar{T}^{(1)}_{K}(y) = 
\underbrace{\frac{1}{\pi}\,\frac{y^2+a^2+1}{\left[\left(y+a\right)^2+1\right]\,\left[\left(y-a\right)^2+1\right]}}_
{W_1(y)} -
\underbrace{\frac{1}{\pi}\,\frac{y^2-a^2+1}{\left[\left(y+a\right)^2+1\right]\,\left[\left(y-a\right)^2+1\right]}}_{W_2(y)}\;,
\end{equation}
\end{widetext}
and also that $\bar{T}^{(1)}_{k}(y)<2\,W_1(y)$. 
An efficient strategy to sample $\bar{T}^{(1)}_{k}(y)$ is, therefore, to sample $W_1(y)$ and then to make use of the usual
rejection technique (see Ref.~\onlinecite{Kalos:book}). 
The primitive of $W_1(y)$ can be easily computed. It reads:
\begin{eqnarray}
M_1(u) &=& \int_{-\infty}^{u}\,{\rm d}\,y\,W_1(y) \nonumber \\
&=&\frac{1}{2\,\pi}\tan^{-1}\left( \frac{2\,u}{1-u^2+a^2}\right)+\frac{1}{2}\;.
\end{eqnarray}
Inverting $M_1(u)$, and making use of an equidistributed random number $u\in(0,1]$, we find that:
\begin{equation}
y =\left\{
\begin{array}{ll}
-d -\sqrt{d^2 + (a^2+1)} & \mbox{for} \; u\in(0,{1\over2}] \\
-d +\sqrt{d^2 + (a^2+1)} & \mbox{for} \; u\in({1\over2},1]
\end{array} \right.
\end{equation}
where
\begin{equation}
d =\cot\left[2\,\pi\,\left(u-{1\over 2}\right) \right]\;.
\end{equation}
Provided $u'\in(0,1]$ -- another equidistributed random number -- we shall accept the $y$ 
obtained above according to the condition:
\begin{equation}
2\,W_1(y)\,u' \leq W_1(y) - W_2(y)\;.
\end{equation}
It results from standard considerations (see Ref.~\onlinecite{Kalos:book})
that the average acceptance of this method will be around $50\%$.
(This means that in order to generate $N$ random numbers distributed according to $T^{(1)}_{K}(x'_i)$,
one has to try on average $2\,N$ times.)
Finally, the original middle-point value $x'_i$ appearing in $T^{(1)}_{K}(x'_i)$
is obtained by inverting the original change of variables in Eq.~(\ref{change:eqn}), 
\begin{equation}
x'_i = \Gamma\,\Delta\,t\,\left(y+c\right)\;.
\end{equation}
This sampling method is a bit cumbersome, but it can be easily implemented as a computer algorithm.
Its generalization, namely
$T^{(l)}_{K}(x'_i)$, which represents the probability of moving the $i$-th bead to $x'_i$ having fixed
the $(i+2^{l-1})$-th and the $(i-2^{l-1})$-th beads in $x_{i+2^{l-1}}$ and
$x_{i-2^{l-1}}$, can be easily derived (it is enough to change $\Delta t$ with $2^{l-1}\,\Delta t$).
This is what one needs in order to implement the bisection scheme \cite{CeperleyPIMC:review}. 

In this way, a PIMC algorithm using such a bisection sampling scheme implements correctly a \emph{relativistic} choice 
of Hamiltonian as in Eq.~\ref{Hrel:eqn}.
Moreover, since $[V,[|p|,V]]=0$, for such a relativistic-PIMC the simple primitive approximation is
already exact to fourth-order \cite{Takahashi_Imada:article}.
We finally report the virial-centroid estimators for the kinetic an
potential energy for the relativistic-PIMC case:
\begin{eqnarray}
K_{PA}  &=& \frac{1}{\,\beta} + {1 \over \,P}\sum_{i=0}^{P-1}\,(x_i - \bar{x})\,\frac{\partial\,V(x_i)}{\partial\,x_i} 
\nonumber \\
V_{PA} &=& {1\over P} \sum_{i=1}^{P}V(x_i)\;, \nonumber
\end{eqnarray}
where $\bar{x}= {1\over P}\,\sum_{i=0}^{P-1}\,x_i$ is the centroid coordinate.
The way to derive them is completely similar to the non-relativistic case
considered in Refs.~\onlinecite{Herman_virial:article,Brualla:article}.


\begin{thebibliography}{41}
\bibitem[{foo({\natexlab{a}})}]{footnote1}
\bibinfo{note}{We made use of a uniform grid of up to $10^3$ points in the
  interval $[-3,+3]$ of the $x$-space, just as in
  Ref.~\onlinecite{Stella_simple:article}. This discretization appears to be
  suitable for the low-lying part of the spectrum we want to study.}
\bibitem[{har()}]{harmonic:note}
\bibinfo{note}{The crossing occurs between the excitation due to the
  well-to-well transition and the first harmonic excitation in the single well.
  At very low temperature the latter is the smallest non-negative eigenvalue of
  $W$ if the former is saturating (e.g. in the box case, when $\sigma \geq
  \sigma_{cr}$). For a theoretical derivation of the equation
  $\Delta_{harm}\propto T^{1/2}$, see
  Ref.~\protect\onlinecite{Stella:phdthesis}.}
\bibitem[{foo({\natexlab{b}})}]{footnote2}
\bibinfo{note}{For $\sigma\to 0$, we have that $B_{\rm eff}\to B = V_0 -
  V(x_{+})\approx 0.9$, the geometrical barrier height.}
\bibitem[{foo({\natexlab{c}})}]{footnote3}
\bibinfo{note}{Notice that, in the Lorentzian case, the annealing exponent with
  $\sigma_{opt}(T)$ is actually a bit smaller than that obtained with the naive
  choice $\sigma\propto T^{1/2}$ (compare Figs.~\ref{FP_vs_MC_SYM_fdt:fig} and
  \ref{FP_vs_MC_SYM_opt:fig}). This small discrepancy is presumably the result
  of the MC implementation, and of a slight ergodicity loss at small $T$.
  Indeed, we have noticed that, for $\tau>10^4$, the system always ends up, at
  the end of the Lorentzian annealing, in the lower minimum, without any
  component on the metastable minimum. Since, as it turns out, the Lorentzian
  $\sigma_{opt}(T)$ is slightly larger than the $\sigma_0\,T^{1/2}$ used in
  Sec.~\protect\ref{CA_intro:sec}, this causes larger fluctuations of
  the results around the lower minimum.}
\bibitem[{foo({\natexlab{d}})}]{footnote4}
\bibinfo{note}{This strategy parallels in a direct way the philosophy behind
  the classical Simulated Annealing technique, where the temperature $T(t)$ is
  also a parameter which is externally driven in the course of a standard
  equilibrium MC simulation.}
\bibitem[{foo({\natexlab{e}})}]{footnote5}
\bibinfo{note}{Since we want to use the smallest possible $T$ (large $\beta$),
  and the computation cost of the PIMC algorithm scales with $P$, the use of a
  fourth-order approximation leads to a sensible speed-up of the QA
  simulations.}
\bibitem[{foo({\natexlab{f}})}]{footnote6}
\bibinfo{note}{We made use of the virial-centroid estimators of the statistical
  functions \protect\cite{Brualla:article,Herman_virial:article}, in order to
  minimize the fluctuations in the averages.}
\bibitem[{foo({\natexlab{g}})}]{footnote7}
\bibinfo{note}{In the definition of $W$, given in
  Sec.~\protect\ref{CA_intro:sec}, we employed $\sigma$ instead of $\Delta t$
  since the first one is the relevant quantity that one can easily tune in a
  practical implementation of the Metropolis algorithm. Nevertheless, it is
  clear that there is a one-to-one correspondence between the move range
  $\sigma$ and the time step $\Delta t$ defined in the above exponential map.}
\end{thebibliography}
%

%
\end{document}